%% file: bsmumuphi_prl_final_he-pex.tex
\documentclass[aps,prl,showpacs,twocolumn,groupedaddress]{revtex4}  % for submission
\usepackage{graphicx}  % needed for figures
\usepackage{dcolumn}   % needed for some tables
\usepackage{bm}        % for math
\usepackage{amssymb}   % for math

\newcommand{\fb}{\ensuremath{\mathrm{fb}^{-1}}}
\newcommand{\bs}{\ensuremath{B_s^0}}
\newcommand{\jp}{\ensuremath{J/\psi}}
\newcommand{\ps}{\ensuremath{\psi(2S)}}
\newcommand{\bsmumuphi}{\ensuremath{\bs \rightarrow \phi  \, \mu^+ \mu^- }}
\newcommand{\bdkstarll}{B_d^0  \rightarrow K^\ast  \, \ell^+ \ell^-}
\newcommand{\bukll}{B^\pm  \rightarrow K^\pm \, \ell^+ \ell^-}
\newcommand{\bsjphi}{\ensuremath{\bs\rightarrow \jp\, \phi}}
\newcommand{\bspsiphi}{\ensuremath{\bs\rightarrow \ps\, \phi}}
\newcommand{\gev}{\mbox{\,\rm Ge\kern -0.1emV}}
\newcommand{\mev}{\mbox{\,\rm Me\kern -0.1emV}}

\begin{document}

%\leftline{Version 1.3 as of \today}
%\leftline{Primary authors: Ralf Bernhard and Frank Lehner}
%\rightline{Comment to {\tt d0-run2eb-023@fnal.gov}}
%\rightline{by 16. March, 2006}
\hspace{5.2in} \mbox{FERMILAB-PUB-06-073-E}
\title{Search for the Rare Decay $\bsmumuphi$ with the D\O~Detector
%\footnote{In this note the charge conjugated states are included
%  implicitly.}
}
\input list_of_authors_r2.tex  % input Dzero author list
\date{\today}

\begin{abstract}
We present a search for the flavor-changing neutral current
decay $\bsmumuphi$ using about 0.45~\fb~of data collected in $p\bar p$ collisions at $\sqrt{s}$=1.96 TeV
with the D\O~detector at the Fermilab Tevatron Collider. We find an upper limit on the branching ratio of this decay normalized to
$\bsjphi$ of $\frac{{\cal B}(\bsmumuphi)}{{\cal B}(\bsjphi)} < 4.4\times 10^{-3}$
at the 95\% C.L. Using the central value of the world average branching fraction of $\bsjphi$, the limit corresponds to ${\cal B}(\bsmumuphi) < 4.1\,\times 10^{-6}$ at the 95\% C.L., the most stringent upper bound
%on the branching
%fraction of $\bsmumuphi$
to date.
\end{abstract}
% activate the following line for publication
\maketitle
The investigation of rare flavor changing neutral current (FCNC) $B$
meson decays has received special attention in the past since
this opens up the possibility of precision
tests of the flavor structure of the standard model (SM). In the SM, FCNC
decays are absent at tree level but proceed at higher order through electroweak penguin
and box diagrams. FCNC decays are sensitive to new physics, since
decay amplitudes involving new particles interfere with SM
amplitudes. Although inclusive FCNC decays like $B\rightarrow X_s\ell^+ \ell^-$ or $B\rightarrow X_s \gamma$ are theoretically easier to
calculate, exclusive decays with one hadron in the final state are
experimentally easier to study. For instance, the exclusive decays $\bdkstarll$ and $\bukll$ have been already measured at $B$-factories~\cite{Aubert:2005cf,Ishikawa:2003cp} and were found to be consistent with the SM within the present experimental uncertainties. Related to the same quark-level transition of $b \rightarrow s\,
\ell^+\ell^-$ is the corresponding exclusive FCNC decay $\bsmumuphi$ in the $B_s^0$ meson system. An observation of this decay or experimental upper limit on its rate will yield additional important information on the flavor dynamics of FCNC decays.

Within the SM, the decay rate for the $\bsmumuphi$ decay, neglecting the interference effects with the much stronger $\bsjphi$ and $\bspsiphi$ resonance decays, is predicted to be of the order of $1.6\times 10^{-6}$~\cite{geng} with about 30\% uncertainty due to
poorly known form factors. The interference effects with the $\bs$ resonance decay amplitudes
are large, with their expected magnitude depending
on the exact modeling of the charmonium states~\cite{polosa}.
To separate experimentally the FCNC-mediated process  $\bsmumuphi$, one has to restrict the invariant mass of the final state lepton pair to be outside the charmonium resonances.
%In the two-Higgs doublet model, the branching fraction of the exclusive
%decay $\bsmumuphi$ might be enhanced~\cite{erkol}, depending on the parameter values of $\tan\beta$ and the mass of the charged Higgs.
Presently, the only existing experimental bound on the
$\bsmumuphi$ decay is given by
CDF from the analysis of Run I data~\cite{CDF}. CDF sets an upper limit at the 95\%
C.L. of ${\cal B}(\bsmumuphi)<6.7\times 10^{-5}$.

In this Letter, we report on a new experimental limit on the decay $\bsmumuphi$, that is an order of magnitude more stringent than the existing limit. The $\phi$ mesons are reconstructed through their $K^+K^-$ decay mode and the invariant mass of the two muons
in the final state is required to be outside the charmonium resonances. The events in
our search are normalized to resonant decay $\bsjphi$ events.
Using the $\bsjphi$ mode as the normalization channel has the advantage that the
efficiencies to detect the $\phi\,\mu^+\mu^-$ system in signal and
normalization events are similar, and systematic effects tend to cancel.

The search uses a data set corresponding to approximately 0.45~\fb of $p\bar{p}$
collisions at $\sqrt{s}=1.96$~TeV recorded by the D\O~detector operating at the Fermilab Tevatron Collider. The D\O \ detector is described in detail elsewhere~\cite{d0nim}. The main elements relevant for this analysis are the central tracking and muon detector systems. The central tracking system consists of a silicon microstrip tracker (SMT) and a central fiber tracker (CFT), both located within a 2~T superconducting
solenoidal magnet. The muon detector, which is located outside the calorimeter,
consists of a layer of tracking detectors and scintillation trigger
counters in front of 1.8~T toroidal magnets, followed by two more similar layers after the toroids, allowing for
efficient muon detection out to pseudorapidity ($\eta)$ of $\pm 2.0$.

Dimuon triggers were used in the data selection for
this analysis. A trigger simulation was used to estimate the trigger
efficiency for the signal and normalization samples.
These efficiencies were also checked with data
samples collected with single muon triggers.
The event pre-selection starts with a loose selection of $\bsmumuphi$ candidates.
These candidates are identified by requiring exactly two muons fulfilling quality cuts on the number of hits in the muon system and the two additional charged particle tracks to form a good vertex. The reconstructed invariant mass of the $\bs$ candidate should be within $4.4 <m_{\phi\,\mu^+\mu^-}<6.2$~GeV/$c^2$.

We then require the invariant mass of the two muons to be within
$0.5<m_{\mu^+\mu^-}<4.4$~GeV/$c^2$. In this mass region, the
$J/\psi(\rightarrow \mu^+\mu^-)$ and $\psi(2S)(\rightarrow \mu^+\mu^-)$
resonances are excluded to discriminate against dominant resonant decays by rejecting the mass region $2.72<m_{\mu^+\mu^-}<4.06$~GeV/$c^2$. The $J/\psi$ mass resolution in data is given by a Gaussian distribution with $\sigma=75$~MeV/$c^2$. The rejected mass region then covers $\pm 5 \sigma$ wide windows around the resonance masses.

The $\chi^2/d.o.f.$ of the two-muon vertex is required to be less than
10. The tracks that are matched to each muon are required to have
at least three (four) measurements in the SMT (CFT) and the transverse
momentum of each of the muons ($p_T^{\mu}$) is required to be greater
than 2.5~GeV/$c$ with $|\eta|<2.0$ to be well inside the fiducial
tracking and muon detector acceptances.
In order to select well-measured secondary vertices,
we define the two-dimensional decay length $L_{xy}$ in the plane
transverse to the beamline, and require its uncertainty $\delta
L_{xy}$ to be less than 0.15~mm. $L_{xy}$ is calculated
as $L_{xy}=\frac{\vec{l}_{vtx}\cdot \vec{p}_T^B}{p_T^B}$, where $p_T^B$ is the
transverse momentum of the candidate $\bs$, and $\vec{l}_{vtx}$ represents
the vector pointing from the primary vertex to the secondary vertex. The uncertainty on the transverse decay
length, $\delta L_{xy}$, is calculated by taking into account the
uncertainties in both the primary and secondary vertex positions. The
primary vertex itself is found for each event using a beam-spot
constrained fit as described in Ref.~\cite{delphi}.

Next, the number of $\bsmumuphi$ candidates is further
reduced by requiring $p_T^B>5~$GeV/$c$ and asking the $\bs$
candidate vertex to have $\chi^2 < 36$ with 5 $d.o.f.$ The two
tracks forming the $\phi$ candidate are further required to have
$p_T>0.7$~GeV/$c$ and their invariant mass within the range $1.008
<m_{\phi} < 1.032~$GeV/$c^2$. The successive cuts and the remaining
candidates surviving each cut are shown in Table~\ref{tab_eff}.
\begin{table}[h]
\small
\begin{center}
\caption{Number of candidate events surviving the cuts in data used in the pre-selection analysis.}
\begin{tabular}{lcc}\hline\hline
   Cut              & Value  & \# candidates \\ % & relative [\%] & rel. MC eff. [\%] \\
\hline
   Good $\bs$ vertex    &                       & 1555320 \\%&      &   \\
   Mass region (GeV/$c^2$) & 0.5 $<m_{\mu^+ \mu^-} <$4.4
                                            &530892  \\%& 10.7 & 87.8 \\
    & excl. $J/\psi$,$\psi(2S)$    &  \\%&  &  \\
   Muon quality &                           & 276875 \\ %& 57.9 & 75.5 \\
   $\chi^2/d.o.f.$ of vertex   & $< 10$     & 127509  \\%& 62.6 & 93.5 \\
   Muon $p_T$ (GeV/$c$)      & $>$ 2.5       & 73555  \\%& 88.1 & 86.1 \\
   Muon $|\eta|$           & $<$ 2.0       & 72350 \\%& 97.0 & 98.5  \\
   Tracking hits    &   CFT$>$ 3, SMT $>$ 2 & 58012 \\%& 73.7 & 90.1 \\
      $\delta L_{xy}$ (mm)&    $<$  0.15   & 54752 \\%& 98.1 &   99.1 \\
   $B_s^0$ candidate $p_T$ (GeV/c) &$>$ 5.0   &54399  \\%& 42.0 & 89.5 \\
   $B_s^0$  $\chi^2 $ vertex &$<$ 36   & 53195 \\%& 42.0 & 89.5 \\
   Kaon  $p_T$ (GeV/$c$) & $>$ 0.7 & 9639 \\
   $\phi$ mass (GeV/$c^2$) & 1.008 $<m_{\phi} <$1.032 & 2602  \\
\hline\hline
\end{tabular}
\label{tab_eff}
\end{center}
\end{table}
We apply the same selection for the resonant $\bsjphi$ candidates except that the invariant mass of the muon
pair is now required to be within $\pm 250$~MeV/$c^2$
of the $J/\psi$ mass.

For the final event selection, we require the candidate events to satisfy
additional criteria. The long lifetime of the $B^0_s$ mesons allows us
to reject the random combinatoric background. For this purpose we use the decay length significance
$L_{xy}/\delta L_{xy}$ as one of the discriminating variables, since it gives better
discriminating power than the transverse decay length alone.
%as large values of $L_{xy}$ may originate due to large uncertainties.

The fragmentation characteristics of the $b$ quark are such that most of its momentum is carried by the $B$ hadron. Thus the number of
extra tracks near the $\bs$ candidate tends to be small. Therefore the second
discriminant is an isolation variable, ${\cal I}$, of the muon and kaon pairs, defined as:
\begin{equation}
   {\cal I}  =
   \frac{|\vec{p}(\phi\,\mu^+\mu^-)|}{|\vec{p}(\phi\,\mu^+\mu^-)|+
   \sum\limits_{{\rm track}\,i \neq B}{ p_i(\Delta {\cal R} < 1)}
   }.\nonumber
\end{equation}
Here, $\sum\limits_{{\rm track}\,i \neq B}{ p_i}$ is the scalar sum over
all tracks excluding the muon and kaon pairs within a cone of $\Delta {\cal R} < 1$
around the momentum vector $\vec{p}(\phi\,\mu^+ \mu^-)$ of the $\bs$ candidate where
$\Delta {\cal R} = \sqrt{(\Delta\phi)^2 + (\Delta\eta)^2 }$.
The final discriminating variable used is the pointing angle $\alpha$,
defined as the angle between the momentum vector
$\vec{p}(\phi\,\mu^+\mu^-)$ of the $\bs$ candidate and the vector $\vec{l}_{vtx}$ between
the primary and secondary vertices. This requirement ensures consistency between the
direction of the decay vertex and the momentum vector of the $\bs$
candidate.

We generate signal Monte Carlo (MC) events for the decay $\bsmumuphi$
using a decay model which includes the NNLO improved Wilson
coefficients~\cite{Ali} for the short-distance part.
The form factors obtained from QCD light-cone sum rules are taken from
Ref.~\cite{ali_form}. These form factors were originally
determined for $B\to K^*$ transitions and were compared with
experimental measurements of the branching fraction ${\cal
  B}(B^0_d\rightarrow K^*\,\ell^+\ell^-)$ in Ref.~\cite{Ali}. Recently, new
form factors for the $\bs \to \phi$ transition, obtained from the light cone QCD sum rules,
were published~\cite{ball}.
The difference between the form factors in Ref.~\cite{Ali} and those in Ref.~\cite{ball}
reaches 20\% for $m_{\mu^+\mu^-} <$ 1~\gev/$c^2$, while elsewhere it
remains well below 10\%.

The analysis is carried out based on signal MC events in the $\bs$ mass region and on data events in regions outside the experimental signal
window defined as $4.51 < m_{\phi\,\mu^+\mu^-} < 6.13$~GeV/$c^2$.
%An optimization based on these discriminating variables was done on
%signal MC events in the $\bs$ mass region $4.51 < M_{\phi\,\mu^+\mu^-} <
%%6.13$~GeV/c$^2$ and on data events in regions outside the signal window, i.e., in the sidebands.
%The whole mass region of interest is shifted downward with respect
%to the world average $\bs$ mass of $m_{B^0_s}=5369.6\pm 2.4$~MeV/c$^2$~\cite{pdg} by 44~MeV/c$^2$ in order to correct for the mass scale of the D\O \ tracker.
A 44~MeV/$c^2$ mass shift in the mass region of interest is
introduced to calibrate the D\O~tracker.
%for the momentum scale of
%taken from the mean $\bs$ mass obtained
%from the fit to the $\bsjphi$ mass spectra without constraining the $\mu^+\mu^-$-pair to the $J/\psi$ mass.

In order to avoid biasing the analysis procedure, data candidates in the signal mass region are not
examined until completion of the analysis, and events in the sideband regions around
the $\bs$ mass are used instead.
The expected mass resolution for $\bsmumuphi$ in the MC is 75~MeV/$c^2$.
The start (end) of the upper (lower) sideband was
chosen such that it is at least 270~MeV/$c^2$ away from
the $\bs$ mass. The widths of the sidebands used
for background estimation are chosen to be  540~MeV/$c^2$ each. The
size of the blind signal region is $\pm 225$~MeV/$c^2$ which
corresponds to a $\pm 3\sigma$ region around the $\bs$
mass. To determine the final limit on the branching fraction, we use a
smaller mass region of $\pm 2.5\sigma$.

A random-grid search~\cite{rgs} was used to find simultaneously the optimal values of the discriminants by maximizing the figure of merit~\cite{punzi} $P=\epsilon_{sig}/(a/2+\sqrt{N_{\rm back}})$.
Here, $\epsilon_{sig}$ is the reconstruction efficiency of the
signal events relative to the preselection (estimated using MC), and
$N_{\rm back}$ is the expected number of background events interpolated
from the sidebands. The constant $a$ is the number of standard deviations
corresponding to the confidence level at which the signal hypothesis
is tested. This constant $a$ was set to 2.0, corresponding to about the 95\% C.L.
After optimization, we find the following
values for the discriminating variables: $L_{xy}/\delta L_{xy}>10.3$, ${\cal I}>0.72$, and $\alpha<0.1$~rad.
%and individual MC signal efficiencies
%including their statistical uncertainties relative to the preselected sample: $L_{xy}/\delta L_{xy}>10.3$
%(73$\pm$4\%), ${\cal I}>0.72$ (92$\pm$5\%), and $\alpha<0.1$~rad (81$\pm$4\%).
%including their statistical uncertainties relative to the preselected sample

The total signal efficiency relative to pre-selection of the three
discriminating cuts is $(54\pm 3)\%$ where the uncertainty is statistical only.
After a linear interpolation of the sideband population for the whole data sample into the mass window signal region, we obtain an
expected number of 1.6$\pm$0.4 background events with statistical
uncertainty only.

\begin{figure}[h]
  \includegraphics[width=\linewidth]{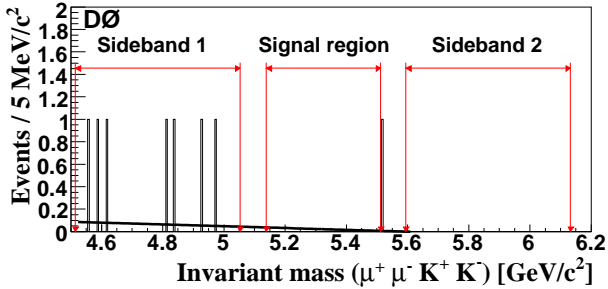}
  \caption{\label{fig_background_standard}The invariant mass
    distribution after optimized
    requirements on the discriminating variables. The solid line shows the sidebands background interpolation into the signal region.}
\end{figure}

Upon examining the data in the mass region, zero candidate events
are observed in the signal region, consistent with the background events as
estimated from sidebands. Figure~\ref{fig_background_standard} shows the
remaining events populating the lower and upper sidebands. The Poisson
probability of observing zero events for an expected background of
$1.6\pm0.4$ is $p=0.22$.

In the absence of an apparent signal, a limit on the
branching fraction ${\cal B}(\bsmumuphi)$ can be computed by normalizing the upper limit on
the number of events in the $B^0_s$ signal region to the number of
reconstructed $\bsjphi$ events:
%In the absence of a signal in our search region we can calculate an
%upper limit on the ratio $(\bsmumuphi/\bsjphi)$ using
\begin{equation}\label{eq_limit2}
\small
\frac{{\cal B}(\bsmumuphi)}{{\cal B}(\bsjphi)} =\frac{N_{\rm ul}}{N_{\bs}}\cdot
 \frac{\epsilon_{J/\psi\phi}}{\epsilon_{\phi\mu^+\mu^-}}\cdot {\cal B}( J/\psi \rightarrow \mu^+ \mu^- ),
\end{equation}
where $N_{\rm ul}$ is the upper limit on the number of signal decays,
estimated from the number of observed events and expected background
events, and $N_{\bs}$ is the observed number of $\bsjphi$ events.
The measured branching fractions are ${\cal B}( J/\psi
\rightarrow \mu^+\mu^-)=(5.88\pm 0.10)\times 10^{-2}$ and ${\cal B}(\bsjphi) = (9.3 \pm 3.3) \times 10^{-4}$~\cite{pdg}. The global efficiencies of the signal and normalization channels are $\epsilon_{\phi\mu^+\mu^-}$ and $\epsilon_{J/\psi\phi}$ respectively, and include all event selection cuts and the acceptance relative to the entire di-muon mass region. They are determined from MC yielding an efficiency ratio of $(\epsilon_{J/\psi \phi}/\epsilon_{\phi\mu^+\mu^-})=2.80\pm 0.21 $, where the uncertainty is due to MC statistics. Applying no cut around the charmonium resonances the efficiency ratio would be $(\epsilon_{J/\psi \phi}/\epsilon_{\phi\mu^+\mu^-}')=1.06\pm 0.07 $.
In order to avoid large uncertainties associated with the poorly known
branching fraction of $\bsjphi$, we normalize the limit of $\bsmumuphi$ relative
to ${\cal B}(\bsjphi)$ as shown by Eq.~\ref{eq_limit2}.

\begin{figure}[h]
  \includegraphics[width=\linewidth]{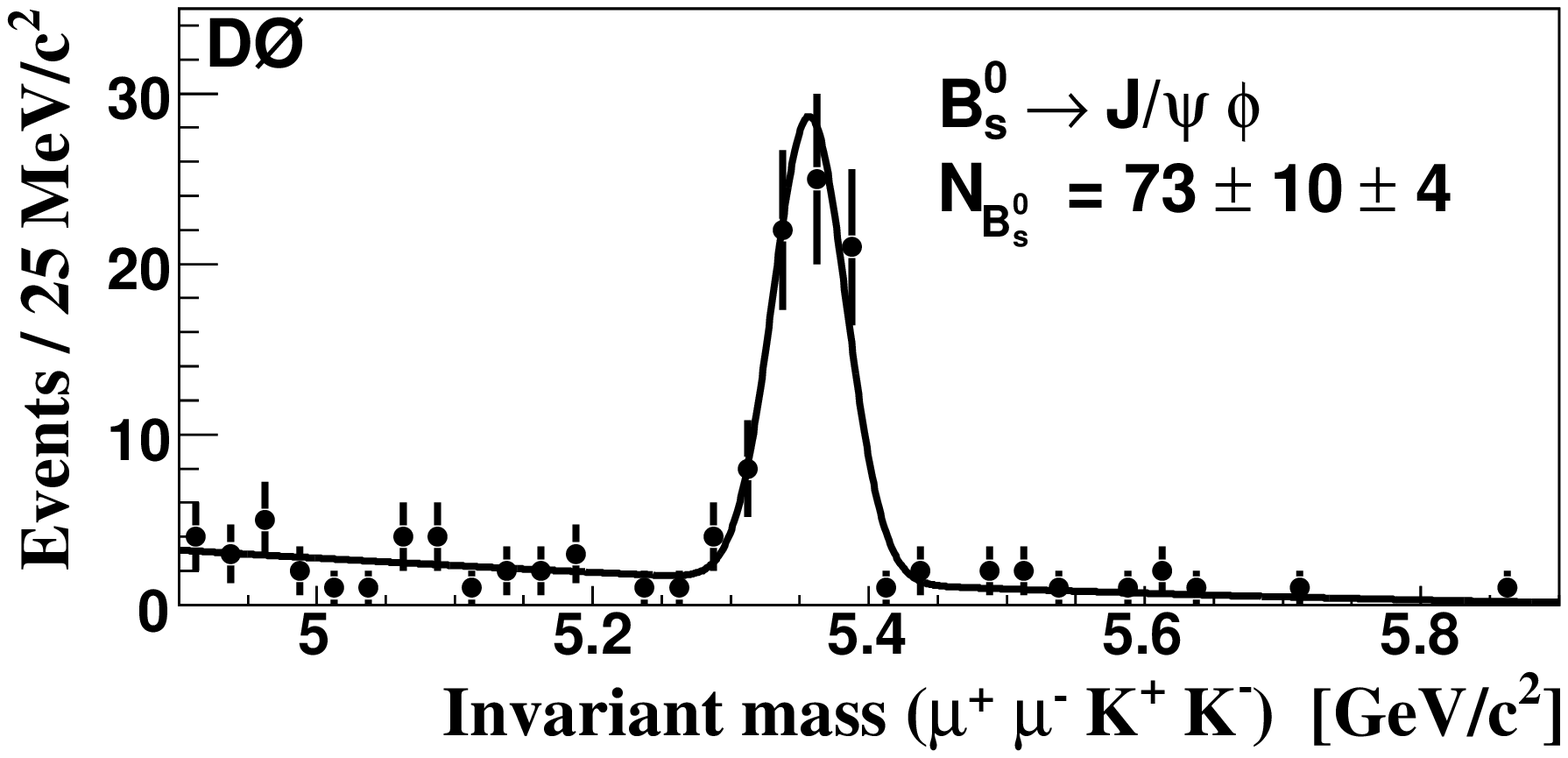}
  \caption{\label{fig_norm}The normalization channel $\bsjphi$.}
\end{figure}

The same cuts are applied to the
$\bsjphi$ candidates. The contamination of muon
pairs from the non-resonant $\phi\,\mu^+\mu^-$ decay in the resonant
normalization region $J/\psi(\rightarrow \mu^+\mu^-)\phi$ is
negligible. We therefore constrain the two muons to have an invariant
mass equal to the $J/\psi$ mass~\cite{pdg} when calculating the $\mu^+\mu^-K^+K^-$
invariant mass.
The mass spectrum of the reconstructed $\bsjphi$ is shown in
Fig.~\ref{fig_norm}. A fit using a Gaussian function for the signal
and a second order polynomial for the background yields $73 \pm 10 \pm
4$ $\bs$ candidates, where the first uncertainty is due to statistics
and the second represents the systematic uncertainty which is estimated by varying the
fit range as well as the background and signal shape hypotheses.

The different sources of relative uncertainty that enter into the limit calculation of ${\cal B}$ of  $\bsmumuphi$ are given in Table~\ref{sys}.
The largest uncertainty, 25\%, is due to the background interpolation into the
signal region and is based on the statistical uncertainty of the fit integral.
The uncertainty on the number of observed $\bsjphi$ events
in the normalization channel is 14.8\%.

The $p_T$ distribution of the $\bs$ in data is on average slightly higher
than that from MC. Therefore, MC events for the signal and normalization channels
have been reweighted accordingly and an additional uncertainty of 3.7\% is applied.
The CP-even signal MC events are generated with a $\bs$ lifetime of 1.44~ps~\cite{Abazov:2004ce}.
To account for a possible efficiency difference related with the shorter lifetime of the CP-even $\bs$, the signal MC events are weighted according to the combined world average CP-even lifetime~\cite{Group(HFAG):2005rb}. The efficiency difference is estimated to be 8\% which is taken as an additional systematic uncertainty.
%The efficiency difference between the CP-even and CP-odd state for the
%normalization channel was estimated using MC to be 6\% with an
%uncertainty on the ratio of 8\%. Therefore another uncertainty
%concerning the CP odd-even eigenstates of the $\bsjphi$ events of 8\% was assigned.
The statistical uncertainty on the efficiency ratio $\epsilon_{J/\psi\phi}/\epsilon_{\phi\mu^+\mu^-}$ is found to be 7.5\%.
The signal efficiency obtained from MC is based on the input for the NNLO Wilson coefficients and form factors of Ref.~\cite{Ali}. We do not include any theoretical uncertainty in our systematics uncertainty estimation.
\begin{center}
\begin{table}[h]
\caption{The relative uncertainties found for the upper limit on ${\cal B}$.}
%Note that the first uncertainty is not taken into
%  account since we normalize the limit to ${\cal B}(\bsjphi)$.}
\begin{center}
  \begin{tabular}{lc}\hline\hline\label{sys}
    Source & Relative Uncertainty $[\%]$ \\ \hline
    $\#$ of $B_s^0 \rightarrow J/\psi \phi$ & 14.8 \\
%    ${\cal B}(B_s^0 \rightarrow J/\psi\, \phi)$ & 35.5 \\\hline
%    $\epsi_{\mu\mu \phi}$ & 5.4 \\
%    $\epsi_{J/\psi \phi}$ & 2.7 \\
    $\epsilon_{J/\psi\phi}/\epsilon_{\phi\mu\mu}$  & 7.5 \\
    MC weighting                            & 3.7 \\ %%sqrt(1.1^2+3.5^2)
%    pre-Geant weighting                     & 3.5 \\
    CP-even lifetime        & 8.0 \\
    ${\cal B}( J/\psi \rightarrow \mu \mu )$ & 1.7 \\    \hline
    Total & 18.9 \\\hline
    Background uncertainty                   & 25.0 \\    \hline \hline
\end{tabular}
\end{center}
\end{table}
\end{center}
The statistical and systematic uncertainties can be included
in the limit calculation by integrating over
probability functions that parameterize the uncertainties. We
use a prescription~\cite{conrad} where we construct a frequentist
confidence interval with the Feldman and Cousins~\cite{feldman}
ordering scheme for the MC integration. The background is modeled as a
Gaussian distribution with its mean value equal to the expected number
of background events and its standard deviation equal to the background
uncertainty. Including the statistical and systematic uncertainties,
the Feldman and Cousins (FC) limit is
\begin{center}
$\frac{{\cal B}(\bsmumuphi)}{{\cal B}(\bsjphi)} <4.4\,(3.5)\,\times 10^{-3}$
\end{center}
at the 95\% (90\%) C.L. respectively. Taking a Bayesian approach~\cite{dzero3476} with a flat prior and the
uncertainties treated as Gaussian distributions in the integration, we find an
upper limit of ${\cal B}(\bsmumuphi)/{\cal B}(\bsjphi) <7.4 \,(5.6)
\times 10^{-3}$ at the 95\% (90\%) C.L., respectively.

Since we have fewer events observed than expected, we also quote the sensitivity of our search.
Assuming there is only background, we calculate for each possible value of
observation a 95\% C.L. upper limit weighted by the Poisson
probability of occurrence. Including the statistical and systematical
uncertainties, our sensitivity is given by $\langle{\cal
  B}(\bsmumuphi)\rangle/{\cal B}(\bsjphi) =1.1\, (1.2) \times 10^{-2}$
at the 95\% C.L. using the FC (Bayesian) approaches, respectively.

Using only the central value of the world average branching fraction~\cite{pdg} of
${\cal B}(\bsjphi) = (9.3 \pm 3.3)\,\times 10^{-4}$, the FC limit
corresponds to ${\cal B}(\bsmumuphi) < 4.1\,(3.2)\,\times 10^{-6}$ at
the 95\% (90\%) C.L. respectively. This is presently the most stringent upper bound and can be compared
%\footnote{A SM calculation~\cite{lunghi} incorporating the actual D\O \ invariant di-muon mass cuts lead to ${\cal B}'(B_s^0\to \phi\,\mu^+\mu^-)=(0.8\pm 0.1)\times 10^{-6}$. This number should then be compared to a limit obtained with the uncorrected efficiency ratio of $\epsilon_{J/\psi\phi}/\epsilon'_{\phi\mu^+\mu^-}=1.06\pm 0.07$.}
with the SM calculation of ${\cal B}(B_s^0\to \phi\,\mu^+\mu^-)=1.6\times 10^{-6}$ of Ref.~\cite{geng}.
%This limit is the most stringent upper bound on the branching fraction $\bsmumuphi$ to date.

\input acknowledgement_paragraph_r2.tex
\end{document}

%% file: list_of_authors_r2.tex
% LIST_OF_AUTHORS_R2.TEX                 3/12/06            
%
\author{                                                                      
%% names begin here                                                           
V.M.~Abazov,$^{36}$                                                           
B.~Abbott,$^{76}$                                                             
M.~Abolins,$^{66}$                                                            
B.S.~Acharya,$^{29}$                                                          
M.~Adams,$^{52}$                                                              
T.~Adams,$^{50}$                                                              
M.~Agelou,$^{18}$                                                             
J.-L.~Agram,$^{19}$                                                           
S.H.~Ahn,$^{31}$                                                              
M.~Ahsan,$^{60}$
% ...
\\                                                                            
\vskip 0.30cm                                                                 
\centerline{(D\O\ Collaboration)}                                             
\vskip 0.30cm                                                                 
}                                                                             
\affiliation{                                                                 
\centerline{$^{1}$Universidad de Buenos Aires, Buenos Aires, Argentina}       
\centerline{$^{2}$LAFEX, Centro Brasileiro de Pesquisas F{\'\i}sicas,         
                  Rio de Janeiro, Brazil}                                     
\centerline{$^{3}$Universidade do Estado do Rio de Janeiro,                   
                  Rio de Janeiro, Brazil}                                     
\centerline{$^{4}$Instituto de F\'{\i}sica Te\'orica, Universidade            
                  Estadual Paulista, S\~ao Paulo, Brazil}                     
\centerline{$^{5}$University of Alberta, Edmonton, Alberta, Canada,           
                  Simon Fraser University, Burnaby, British Columbia, Canada,}
\centerline{York University, Toronto, Ontario, Canada, and                    
                  McGill University, Montreal, Quebec, Canada}                
\centerline{$^{6}$Institute of High Energy Physics, Beijing,                  
                  People's Republic of China}                                 
\centerline{$^{7}$University of Science and Technology of China, Hefei,       
                  People's Republic of China}                                 
\centerline{$^{8}$Universidad de los Andes, Bogot\'{a}, Colombia}             
\centerline{$^{9}$Center for Particle Physics, Charles University,            
                  Prague, Czech Republic}                                     
\centerline{$^{10}$Czech Technical University, Prague, Czech Republic}        
\centerline{$^{11}$Center for Particle Physics, Institute of Physics,         
                   Academy of Sciences of the Czech Republic,                 
                   Prague, Czech Republic}                                    
\centerline{$^{12}$Universidad San Francisco de Quito, Quito, Ecuador}        
\centerline{$^{13}$Laboratoire de Physique Corpusculaire, IN2P3-CNRS,         
                   Universit\'e Blaise Pascal, Clermont-Ferrand, France}      
\centerline{$^{14}$Laboratoire de Physique Subatomique et de Cosmologie,      
                   IN2P3-CNRS, Universite de Grenoble 1, Grenoble, France}    
\centerline{$^{15}$CPPM, IN2P3-CNRS, Universit\'e de la M\'editerran\'ee,     
                   Marseille, France}                                         
\centerline{$^{16}$IN2P3-CNRS, Laboratoire de l'Acc\'el\'erateur              
                   Lin\'eaire, Orsay, France}                                 
\centerline{$^{17}$LPNHE, IN2P3-CNRS, Universit\'es Paris VI and VII,         
                   Paris, France}                                             
\centerline{$^{18}$DAPNIA/Service de Physique des Particules, CEA, Saclay,    
                   France}                                                    
\centerline{$^{19}$IReS, IN2P3-CNRS, Universit\'e Louis Pasteur, Strasbourg,  
                    France, and Universit\'e de Haute Alsace,                 
                    Mulhouse, France}                                         
\centerline{$^{20}$Institut de Physique Nucl\'eaire de Lyon, IN2P3-CNRS,      
                   Universit\'e Claude Bernard, Villeurbanne, France}         
\centerline{$^{21}$III. Physikalisches Institut A, RWTH Aachen,               
                   Aachen, Germany}                                           
\centerline{$^{22}$Physikalisches Institut, Universit{\"a}t Bonn,             
                   Bonn, Germany}                                             
\centerline{$^{23}$Physikalisches Institut, Universit{\"a}t Freiburg,         
                   Freiburg, Germany}                                         
\centerline{$^{24}$Institut f{\"u}r Physik, Universit{\"a}t Mainz,            
                   Mainz, Germany}                                            
\centerline{$^{25}$Ludwig-Maximilians-Universit{\"a}t M{\"u}nchen,            
                   M{\"u}nchen, Germany}                                      
\centerline{$^{26}$Fachbereich Physik, University of Wuppertal,               
                   Wuppertal, Germany}                                        
\centerline{$^{27}$Panjab University, Chandigarh, India}                      
\centerline{$^{28}$Delhi University, Delhi, India}                            
\centerline{$^{29}$Tata Institute of Fundamental Research, Mumbai, India}     
\centerline{$^{30}$University College Dublin, Dublin, Ireland}                
\centerline{$^{31}$Korea Detector Laboratory, Korea University,               
                   Seoul, Korea}                                              
\centerline{$^{32}$SungKyunKwan University, Suwon, Korea}                     
\centerline{$^{33}$CINVESTAV, Mexico City, Mexico}                            
\centerline{$^{34}$FOM-Institute NIKHEF and University of                     
                   Amsterdam/NIKHEF, Amsterdam, The Netherlands}              
\centerline{$^{35}$Radboud University Nijmegen/NIKHEF, Nijmegen, The          
                  Netherlands}                                                
\centerline{$^{36}$Joint Institute for Nuclear Research, Dubna, Russia}       
\centerline{$^{37}$Institute for Theoretical and Experimental Physics,        
                   Moscow, Russia}                                            
\centerline{$^{38}$Moscow State University, Moscow, Russia}                   
\centerline{$^{39}$Institute for High Energy Physics, Protvino, Russia}       
\centerline{$^{40}$Petersburg Nuclear Physics Institute,                      
                   St. Petersburg, Russia}                                    
\centerline{$^{41}$Lund University, Lund, Sweden, Royal Institute of          
                   Technology and Stockholm University, Stockholm,            
                   Sweden, and}                                               
\centerline{Uppsala University, Uppsala, Sweden}                              
\centerline{$^{42}$Physik Institut der Universit{\"a}t Z{\"u}rich,            
                   Z{\"u}rich, Switzerland}                                   
\centerline{$^{43}$Lancaster University, Lancaster, United Kingdom}           
\centerline{$^{44}$Imperial College, London, United Kingdom}                  
\centerline{$^{45}$University of Manchester, Manchester, United Kingdom}      
\centerline{$^{46}$University of Arizona, Tucson, Arizona 85721, USA}         
\centerline{$^{47}$Lawrence Berkeley National Laboratory and University of    
                   California, Berkeley, California 94720, USA}               
\centerline{$^{48}$California State University, Fresno, California 93740, USA}
\centerline{$^{49}$University of California, Riverside, California 92521, USA}
\centerline{$^{50}$Florida State University, Tallahassee, Florida 32306, USA} 
\centerline{$^{51}$Fermi National Accelerator Laboratory,                     
            Batavia, Illinois 60510, USA}                                     
\centerline{$^{52}$University of Illinois at Chicago,                         
            Chicago, Illinois 60607, USA}                                     
\centerline{$^{53}$Northern Illinois University, DeKalb, Illinois 60115, USA} 
\centerline{$^{54}$Northwestern University, Evanston, Illinois 60208, USA}    
\centerline{$^{55}$Indiana University, Bloomington, Indiana 47405, USA}       
\centerline{$^{56}$University of Notre Dame, Notre Dame, Indiana 46556, USA}  
\centerline{$^{57}$Purdue University Calumet, Hammond, Indiana 46323, USA}    
\centerline{$^{58}$Iowa State University, Ames, Iowa 50011, USA}              
\centerline{$^{59}$University of Kansas, Lawrence, Kansas 66045, USA}         
\centerline{$^{60}$Kansas State University, Manhattan, Kansas 66506, USA}     
\centerline{$^{61}$Louisiana Tech University, Ruston, Louisiana 71272, USA}   
\centerline{$^{62}$University of Maryland, College Park, Maryland 20742, USA} 
\centerline{$^{63}$Boston University, Boston, Massachusetts 02215, USA}       
\centerline{$^{64}$Northeastern University, Boston, Massachusetts 02115, USA} 
\centerline{$^{65}$University of Michigan, Ann Arbor, Michigan 48109, USA}    
\centerline{$^{66}$Michigan State University,                                 
            East Lansing, Michigan 48824, USA}                                
\centerline{$^{67}$University of Mississippi,                                 
            University, Mississippi 38677, USA}                               
\centerline{$^{68}$University of Nebraska, Lincoln, Nebraska 68588, USA}      
\centerline{$^{69}$Princeton University, Princeton, New Jersey 08544, USA}    
\centerline{$^{70}$State University of New York, Buffalo, New York 14260, USA}
\centerline{$^{71}$Columbia University, New York, New York 10027, USA}        
\centerline{$^{72}$University of Rochester, Rochester, New York 14627, USA}   
\centerline{$^{73}$State University of New York,                              
            Stony Brook, New York 11794, USA}                                 
\centerline{$^{74}$Brookhaven National Laboratory, Upton, New York 11973, USA}
\centerline{$^{75}$Langston University, Langston, Oklahoma 73050, USA}        
\centerline{$^{76}$University of Oklahoma, Norman, Oklahoma 73019, USA}       
\centerline{$^{77}$Oklahoma State University, Stillwater, Oklahoma 74078, USA}
\centerline{$^{78}$Brown University, Providence, Rhode Island 02912, USA}     
\centerline{$^{79}$University of Texas, Arlington, Texas 76019, USA}          
\centerline{$^{80}$Southern Methodist University, Dallas, Texas 75275, USA}   
\centerline{$^{81}$Rice University, Houston, Texas 77005, USA}                
\centerline{$^{82}$University of Virginia, Charlottesville,                   
            Virginia 22901, USA}                                              
\centerline{$^{83}$University of Washington, Seattle, Washington 98195, USA}  
}                                                                             
%end                                                                          

%% file: acknowledgement_paragraph_r2.tex
% acknowledgement_paragraph_r2.tex                3/12/06
%
We thank the staffs at Fermilab and collaborating institutions, 
and acknowledge support from the 
DOE and NSF (USA);
CEA and CNRS/IN2P3 (France);
FASI, Rosatom and RFBR (Russia);
CAPES, CNPq, FAPERJ, FAPESP and FUNDUNESP (Brazil);
DAE and DST (India);
Colciencias (Colombia);
CONACyT (Mexico);
KRF and KOSEF (Korea);
CONICET and UBACyT (Argentina);
FOM (The Netherlands);
PPARC (United Kingdom);
MSMT (Czech Republic);
CRC Program, CFI, NSERC and WestGrid Project (Canada);
BMBF and DFG (Germany);
SFI (Ireland);
The Swedish Research Council (Sweden);
Research Corporation;
Alexander von Humboldt Foundation;
and the Marie Curie Program.